\documentclass{article}

\usepackage{url}
\usepackage{algorithm}
\usepackage{algorithmic}
\usepackage{graphicx}
\usepackage{amsmath}
\newcommand{\xRightarrow}[2][]{\ext@arrow 0359\Rightarrowfill@{#1}{#2}}
\usepackage{amsfonts}
\usepackage{multirow}
\usepackage{bm}

\newtheorem{theorem}{Theorem}
\newcommand{\mat}[1]{\bm{#1}}
\newcommand{\ten}[1]{\bm{\mathcal{#1}}} 
\usepackage{subfigure}
\usepackage{xcolor}

\usepackage{ulem}
\usepackage{microtype}
\usepackage{subfigure}
\usepackage{lipsum}
\usepackage{booktabs} % for professional tables
\newcommand\blfootnote[1]{%
\begingroup
\renewcommand\thefootnote{}\footnote{#1}%
\addtocounter{footnote}{-1}%
\endgroup
}
\usepackage{hyperref}

\usepackage[accepted]{icml2021}

\icmltitlerunning{EZCrop: Energy-Zoned Channels for Robust Output Pruning}

\begin{document}

\twocolumn[
\icmltitle{EZCrop: Energy-Zoned Channels for Robust Output Pruning}



\icmlsetsymbol{equal}{*}

\begin{icmlauthorlist}
\icmlauthor{Rui Lin}{equal,to}
\icmlauthor{Jie Ran}{equal,to}
\icmlauthor{Dongpeng Wang}{UMEC}
\icmlauthor{King Hung Chiu}{UMEC}
\icmlauthor{Ngai Wong}{to}
\end{icmlauthorlist}

\icmlaffiliation{to}{Department of Computation, University of Torontoland, Torontoland, Canada}

\icmlcorrespondingauthor{Cieua Vvvvv}{c.vvvvv@googol.com}
\icmlcorrespondingauthor{Eee Pppp}{ep@eden.co.uk}

\icmlkeywords{Machine Learning, ICML}

\vskip 0.3in
]




\blfootnote{$^*$ Equal contribution $^1$ Department of Electrical and Electronic Engineering, The University of Hong Kong. Email adress: \{linrui, jran, nwong\} @eee.hku.hk. $^2$ UMEC - United Microelectronics Centre (Hong Kong) Limited.}

%%%%%%%%% ABSTRACT
\begin{abstract}
Recent results have revealed an interesting observation in a trained convolutional neural network (CNN), namely, the rank of a feature map channel matrix remains surprisingly constant despite the input images. This has led to an effective rank-based channel pruning algorithm~\cite{lin2020hrank}, yet the constant rank phenomenon remains mysterious and unexplained. This work aims at demystifying and interpreting such rank behavior from a frequency-domain perspective, which as a bonus suggests an extremely efficient Fast Fourier Transform (FFT)-based metric for measuring channel importance without explicitly computing its rank. We achieve remarkable CNN channel pruning based on this analytically sound and computationally efficient metric, and adopt it for repetitive pruning to demonstrate robustness via our scheme named \textbf{E}nergy-\textbf{Z}oned \textbf{C}hannels for \textbf{R}obust \textbf{O}utput \textbf{P}runing (EZCrop), which shows consistently better results than other state-of-the-art channel pruning methods. %The source codes of EZCrop will be released upon acceptance of the paper. %
\end{abstract}

%%%%%%%%% BODY TEXT
%%%% Intro
\section{Introduction}
\label{sec:intro}
Convolutional neural networks (CNNs) are among the most popular models for deep learning which have achieved breakthroughs in vision applications including classification~\cite{sun2020automatically}, object detection~\cite{liu2020cbnet}, semantic segmentation~\cite{luo2019taking}, etc. However, deeper and larger CNNs render them challenging to deploy on edge devices with constrained hardware resources. This dilemma has motivated the research on compressing CNNs for lower storage and faster inference. Numerous neural network compression techniques have been proposed, including optimized implementation~\cite{jiang2018efficient}, quantization~\cite{cheng2017quantized, ding2019regularizing}, network pruning~\cite{he2019filter}, low-rank decomposition~\cite{lin2018holistic,kim2020bayesian} and knowledge distillation~\cite{hinton2015distilling,luo2016face}, etc. Among the categories mentioned above, this work focuses on channel pruning, a subclass of structured network pruning. 

Due to its outstanding performance in model size reduction and network acceleration, network pruning has been a popular compression strategy. Generally, it can be divided into two subclasses: weight pruning and filter pruning. Weight pruning~\cite{han2015deep,han2015learning,guo2016dynamic} tends to eliminate small weights in the kernel tensors, leading to a sparse architecture with fewer nonzero connections. However, it is not easy to achieve inference acceleration via the existing efficient BLAS libraries due to the unstructured sparsity pattern, whose utilization requires specialized software or hardware. On the other hand, channel pruning~\cite{li2016pruning,liu2017learning,lin2020hrank} removes entire filters in the kernel tensor. Compared with weight pruning, eliminating entire filters results in structured sparsity and generic speedup irrespective of the software/hardware, which promotes the usefulness of filter pruning.

The key challenge in channel pruning is to evaluate the importance of filters and select trivial or unimportant candidates for pruning. For instance, in~\cite{molchanov2016pruning}, the authors propose a novel criterion based on Taylor expansion to evaluate the filters' importance and discard the less important ones. In~\cite{liu2018frequency}, the authors use 2D discrete cosine transform (DCT) on the filters to exploit the spatial correlations in frequency domain and prune filters containing less low-frequency components dynamically. In~\cite{he2019filter} ``smaller-norm-less-important'' filter pruning is proposed, which removes filters with smaller norms. Recently, Lin~\textit{et al.}~\cite{lin2020hrank} observe that the average rank of feature maps generated by a fixed filter remains surprisingly constant irrespective of the input images. They then propose a scheme called \textit{HRank} to prune channels with small corresponding ranks and thereby less information content.

\begin{figure}[t]
\centering
\includegraphics[scale=0.45]{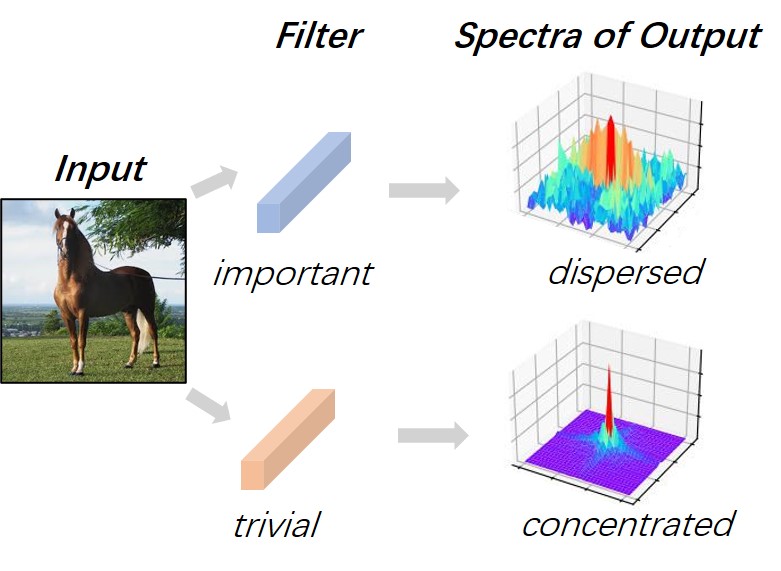}
\caption{The spectral of output is obtained by applying 2D-FFT and ${\rm fftshift}(\cdot)$ on the output in the spatial domain. (Upper) The output of an important filter in the frequency domain has a dispersed energy distribution. (Lower) The output of a trivial filter has a concentrated energy distribution in the frequency domain.}
\label{fig:idea}
\end{figure}

Noticeably, most existing pruning schemes operate in the spatial domain, with only a few of them utilizing information from the frequency or spectral domain. Inspired by the work of Sedghi \textit{et al.}~\cite{sedghi2018singular} that studies the singular values in a convolution layer from a Fast Fourier Transformation (FFT) viewpoint, we take FFT as a tool for deciphering the constant rank behavior~\cite{lin2020hrank}, and develop a filter pruning scheme with a novel, analytical, energy-based explanation. Our key insight is that the energy distributions of feature maps generated by a fixed filter can reflect the filter's importance (Figure~\ref{fig:idea}). Alongside this observation, we propose a novel frequency-domain channel pruning approach called Energy-Zoned Channel for robust output pruning (EZCrop) , which entails a natural, accurate, and more robust selection of important filters at a lower computational cost than the singular value decomposition (SVD) needed for evaluating ranks in HRank~\cite{lin2020hrank}.

In order to make a comprehensive comparison versus HRank~\cite{lin2020hrank} and other state-of-the-art algorithms
~\cite{li2016pruning,hu2016network,liu2017learning,luo2017thinet,yu2018nisp,huang2018data,lin2019towards,he2019filter,wang2020pruning}
, we benchmark EZCrop for image classification tasks on CIFAR-10~\cite{krizhevsky2009learning} and ImageNet~\cite{deng2009imagenet}, using VGGNet~\cite{simonyan2014very}, ResNet~\cite{he2016deep}, and DenseNet~\cite{huang2017densely}. The experimental results demonstrate the \textit{consistent superiority} of EZCrop in both model reduction and acceleration. Moreover, we conduct repetitive pruning with HRank and EZCrop, which further illustrates the high robustness of EZCrop. In summary, our main contributions are:
\begin{itemize}
\item We analytically bridge the rank-based channel importance metric in the spatial domain to an energy perspective in the frequency domain, and for the first time explain the interesting constant-rank phenomenon in a channel matrix.
\item We propose a computationally efficient FFT-based metric for channel importance, which reduces the computational complexity from $\mathcal{O}(n^3)$ to $\mathcal{O}(n^2 {\rm log} n)$.
\item The proposed EZCrop algorithm for channel pruning is simple, intuitive and robust. It outperforms other state-of-the-art pruning schemes and consistently delivers good performance, which is not a result of normal variations as confirmed through extensive experiments.
\end{itemize}

% fig

%%%% Lit review / related work
\section{Related Work}
\label{sec:lit_review}
%% filter pruning
\subsection{Filter Pruning}
\label{sec:lit_pruning}
In contrast to weight pruning that results in unstructured sparsity, channel or filter pruning belongs to the class of structured pruning and can readily leverage the efficient BLAS library without specific software or hardware requirements. Generally, there are two categories of CNN filter pruning schemes, namely, by utilizing post-training properties of the CNN itself or by adaptive pruning during training. The first category employs algebraic metrics. For example, in~\cite{li2016pruning}, the authors use $l_1$-norm of the filters or the corresponding feature maps to evaluate the importance, where smaller norms mean less informative channels. In HRank~\cite{lin2020hrank}, filters with small corresponding average ranks calculated by a batch of feature map slices are pruned. For the second category, adaptive pruning approaches make decisions through retraining with a specialized loss function taking pruning into account. For example, Liu \textit{et al.}~\cite{liu2017learning} impose sparsity-induced regularization on scaling factors in the batch normalization layers, where channels with zero-valued scaling factors are pruned. By reformulating the batch normalization layer,  the authors of \cite{zhao2019variational} propose a new parameter called saliency, and only channels with saliency beyond a threshold during training can remain. 

%% FFT Application   
\subsection{Application of Frequency Information in a CNN}
\label{sec:lit_freq}
In recent years,  the properties of CNNs in the frequency domain have attracted increasing attention due to the acceleration of CONV calculations and the additional information arising from the frequency domain. In~\cite{rippel2015spectral}, the authors propose spectral pooling that preserves more information and complex-coefficient spectral parameterization of filters that improves training convergence. Pratt \textit{et al.}~\cite{pratt2017fcnn} propose a frequency-domain CNN, whose training can be done entirely in the frequency-domain without alternating between spatial and spectral domains. In~\cite{sedghi2018singular}, by using FFT properties, the authors design a regularizer to constrain the range of CONV kernels' singular values during model training, which improves the model performance. Including~\cite{liu2018frequency}, there are other works using spectral information for network compression. For example, in~\cite{chen2016compressing}, the model is compressed in a frequency-sensitive fashion which preserves parameters of low-frequency components better. The filters are treated as images in~\cite{wang2016cnnpack}, where the authors compress the model by discarding the low-energy coefficients of filters in the frequency domain. 

%% Motivation
Summarizing over the above works on filter pruning and frequency-domain CNN design, we notice the research on filter pruning through exploiting frequency-domain information is insufficient. The method in~\cite{liu2018frequency}, though doing dynamic pruning in the frequency domain, belongs to unstructured and adaptive pruning, and suffers from machine and labor costs. Therefore, further exploring the use of spectral information for filter pruning is desirable.

% fig: 3D plot
\begin{figure*}
\centering
\includegraphics[scale=0.40]{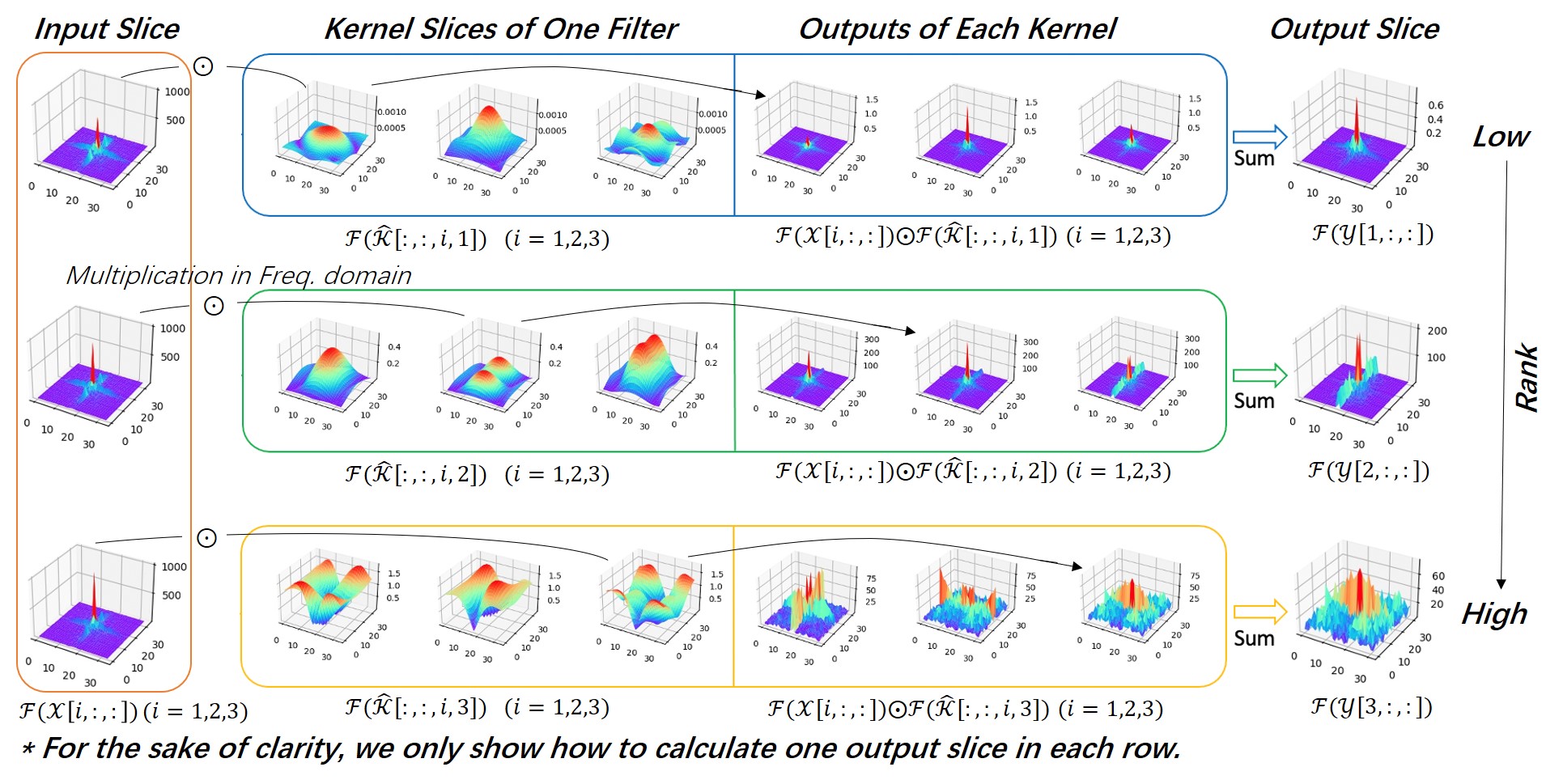}
\caption{The first column shows the three input channels in the frequency domain with center shifting, namely, ${\rm fftshift}(\mathcal{F}(\ten{X}[i,:,:]))$ for $i=1,2,3$. On the left of the second column are expended kernel slices ${\rm fftshift}(\mathcal{F}(\hat{\ten{K}}[:,:,i,j]))$, where $i=1,2,3$ and $j$ is fixed but different in three rows. The corresponding outputs slices ${\rm fftshift}(\mathcal{F}(\hat{\ten{K}}[:,:,i,j])\odot \mathcal{F}(\ten{X}[i,:,:]))$ are shown on the right part, and $\mathcal{F}(\ten{Y}[j,:,:])$ are the sum of three element-wise product, respectively. The selected filters generate feature maps with increasing average ranks from top to bottom. Note the spectral energy distribution of the feature map becomes more dispersed for higher rank spatial filters.}
\label{fig:fftkx}
\end{figure*}

% fig: matrices
\begin{figure}[t]
\centering
\includegraphics[scale=0.37]{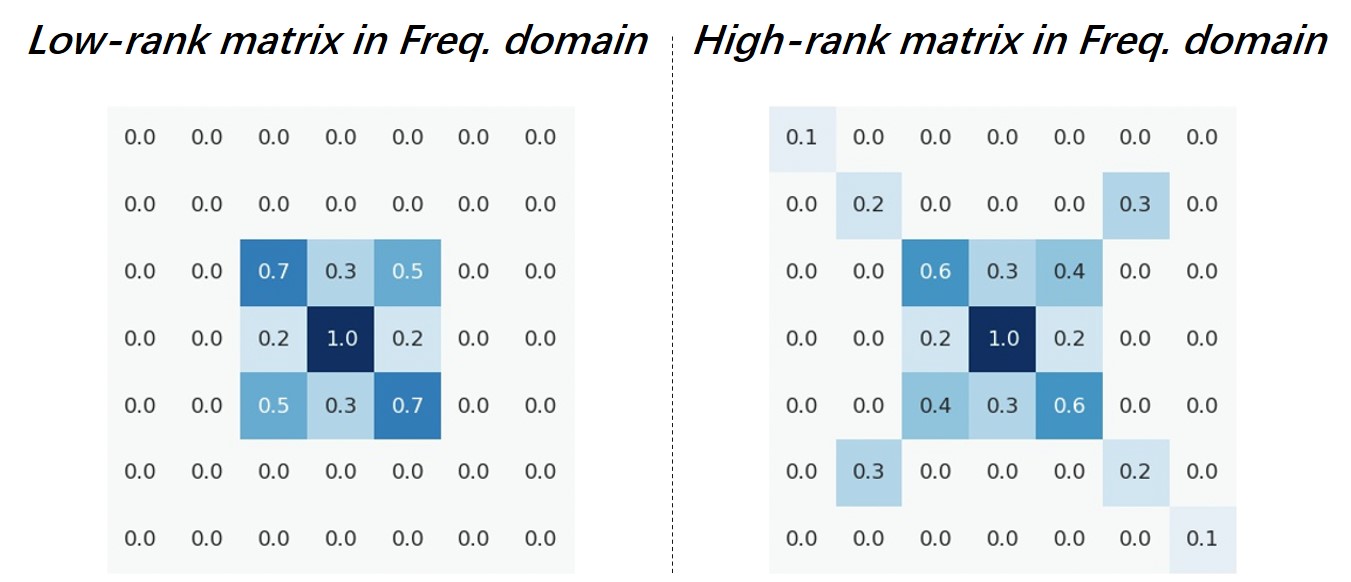}
\caption{Connection between ranks and energy dispersion. When transforming the matrix to the frequency domain, the ${\rm fftshift}(\cdot)$ operation is employed, which is essentially a block permutation and is rank-preserving. Here it is apparent the left matrix is only of rank 3 while the right is full-rank, and such spectral ranks also translate to spatial ranks due to rank-invariant domain transforms.}
\label{fig:rank_energy}
\end{figure}

%%%% spatial --> frequency
\section{Constant Ranks in Feature Map: A Frequency-Domain Perspective}
\label{sec:explain}
The idea proposed in \cite{sedghi2018singular} by using FFT to constrain the singular values of kernels inspires us to use FFT to relate the rank-based metric in the spatial domain to an energy perspective in the frequency domain.

%%%% notation
\subsection{Notation}
\label{sec:notation}
For a trained CNN model, we use $\ten{Y}_i \in \mathbb{R}^{B\times T_i \times H_i \times W_i}$ to denote the output\footnote{The output selected here is after the operations like ReLU, batch normalization, and max pooling, etc. In other words, the output we extract for the $i$-th CONV layer is the input for the $(i+1)$-th CONV layer.} of the $i$-th CONV layer $(i=1,2,\cdots,N)$, where $B$ is the batch size, $T_i$ is the number of channels, $H_i$ and $W_i$ are the height and width, respectively. We denote the kernels of the $i$-th CONV layer as $\ten{K}_i \in \mathbb{R}^{D_i \times D_i \times S_i \times T_i}$, where $D_i$ represents the kernel size, $S_i$ and $T_i$ are the number of input and output channels. We use $\mat{Y}_i^{j}[b,:,:] \in \mathbb{R}^{H_i \times W_i}$ to denote the feature map generated by the $b$-th sample in the batch passing through the $j$-th filter in the $i$-th CONV layer. The set $I_i = \{ I_i^1, I_i^2, \cdots, I_i^{T_i} \}$ is used to record the importance of filters in the $i$-th CONV layer. Every element in the set $I_i$ records the index of one filter. We use $\mathcal{F}(\cdot)$ and $\mathcal{F}^{-1}(\cdot)$ to denote FFT and inverse FFT, respectively. The notation ${\rm fftshift}(\cdot)$ stands for the operation swapping four quadrants to center the DC component, ${\rm abs}(\cdot)$ means element-wise magnitude, and ${\rm ceil}(\cdot)$ means rounding up.

%%%% fig: circle plots
\begin{figure}[t]
\centering
\includegraphics[scale=0.37]{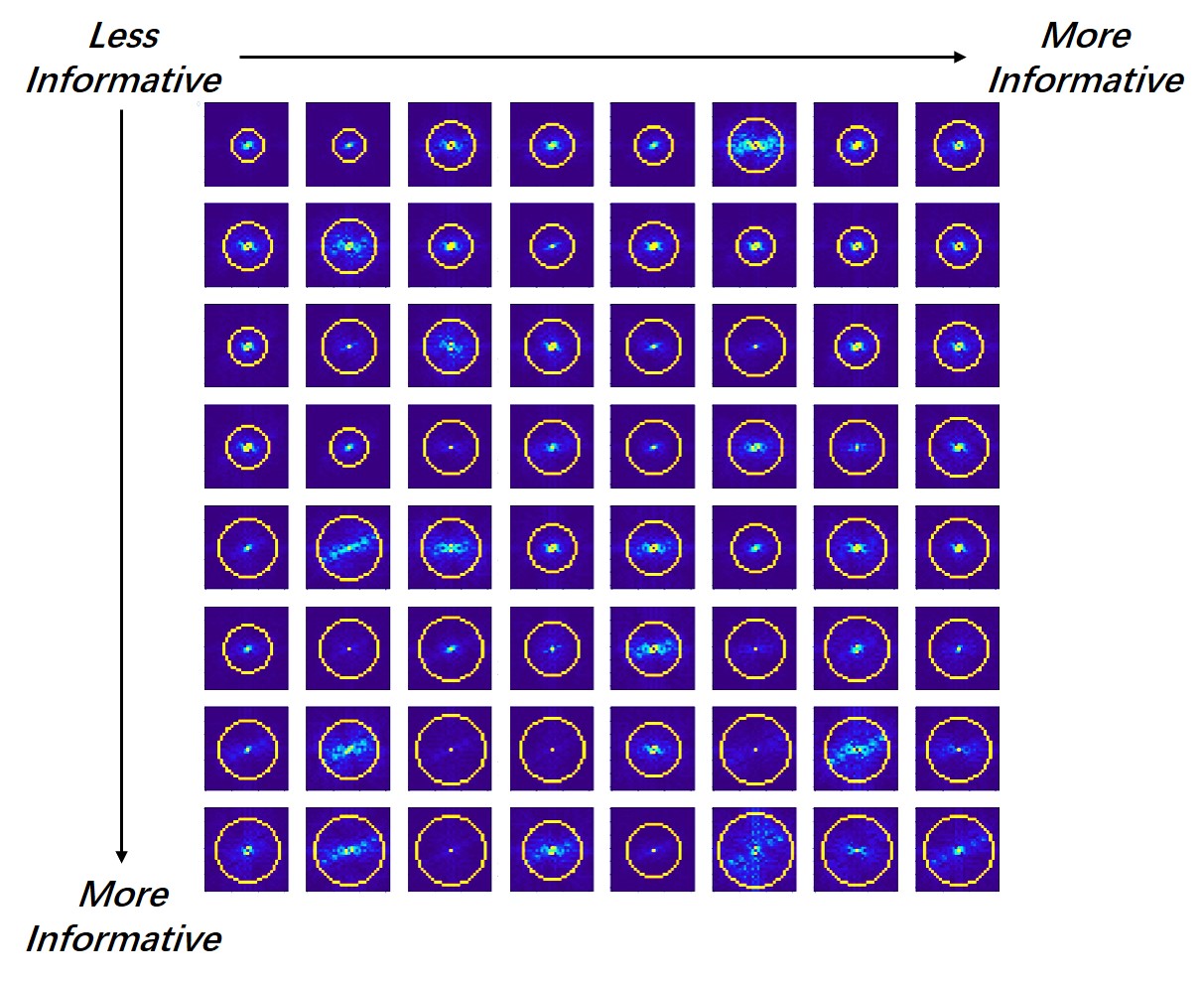}
\caption{The outputs are generated by 64 filters from the first CONV layer of a pretrained VGGNet. The 64 subfigures are arranged in an ascending order of the rank-based HRank scheme from left to right and top to bottom. We determine the radii of the circles to measure and visualize the concentration of the low-frequency components of the feature map in the frequency domain.}
\label{fig:rank_in_frequency1}
\end{figure}

%% fig: Hi and Wi
\begin{figure}[t]
\centering
\includegraphics[scale=0.45]{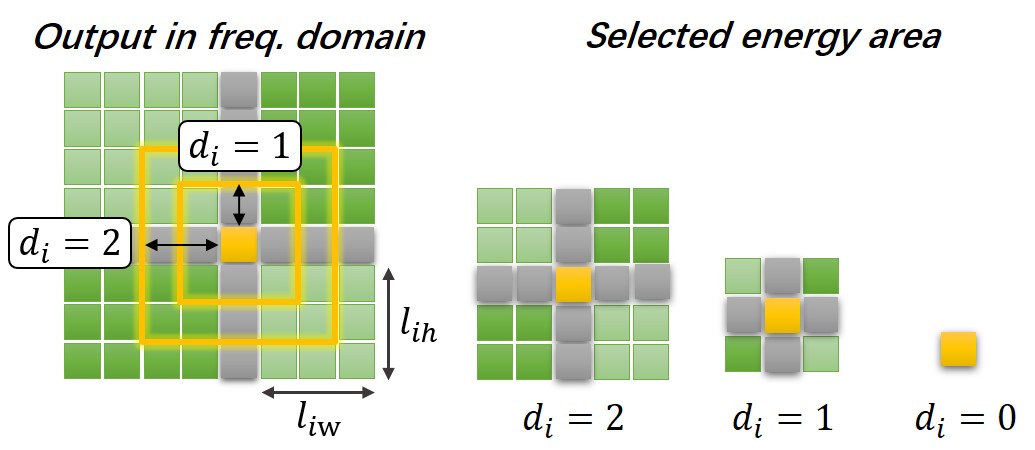}
\caption{An example to illustrate how to find the square center, and determine the selected energy area.}
\label{fig:energy_area}
\end{figure}

%%%% fig: curves
\begin{figure}[t]
\centering
\includegraphics[scale=0.25]{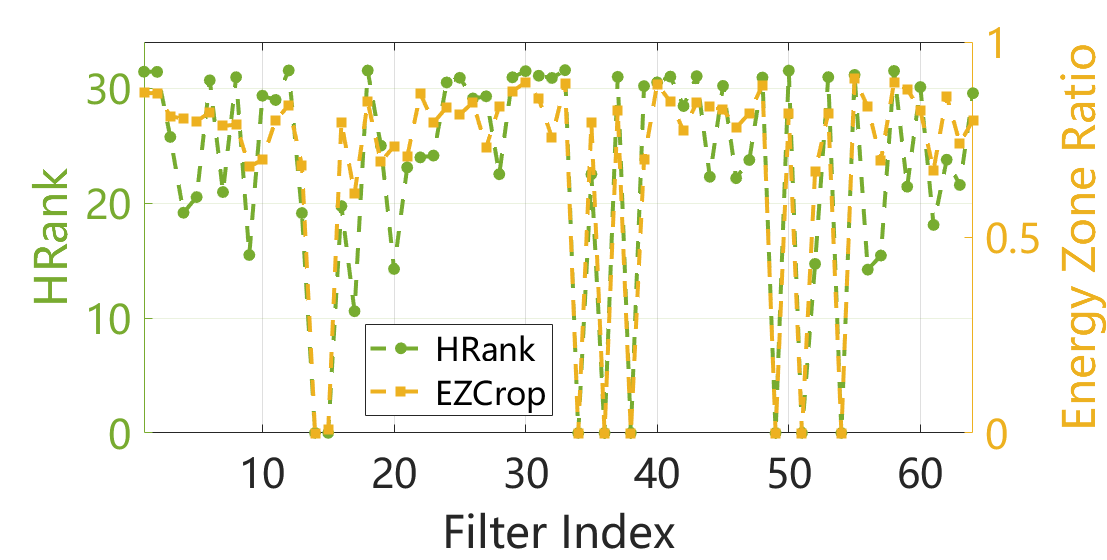}
\caption{The two curves indicate the importance of the $64$ filters using HRank and our energy-based metrics (Eq.~\ref{eq:ratio}) in the spatial and spectral domains, respectively. The dashed lines joining the points are only for the ease of visualization wherein the points are discrete in nature.}
\label{fig:rank_in_frequency2}
\end{figure}

%%%% linkage: convolution --> multiplication
\subsection{Convolution in the Frequency Domain}
\label{sec:connection}
It is well-known that convolution in the time/spatial domain is equivalent to multiplication in the frequency domain. In~\cite{podlozhnyuk2007fft}, the author illustrates how to transform a general $2$D convolution into a Fourier-based one. In brief, there are three steps: 1) for the given convolution kernel and input data, we apply $2$D FFT on them respectively, 2) then we do point-wise multiplication between the newly obtained kernel and input, 3) lastly, we perform inverse $2$D FFT on the multiplied result. Theorem~\ref{theorem:freq_conv} formalizes the convolution process in the frequency domain, and the detailed proof is provided in Appendix 1.

% theorem to show the convolution in Freq. domain
\begin{theorem}
\label{theorem:freq_conv}
For a single $3$-D input $\ten{X} \in \mathbb{R}^{S\times H\times W}$ and a given filter $\ten{K} \in \mathbb{R}^{D\times D\times S \times T}$, their convolution result $\ten{Y} \in \mathbb{R}^{T\times H\times W}$ can be formalized as:
% eq
\begin{equation}
\label{eq:freq_conv}
\ten{Y}[j,:,:] = \sum_{i=1}^{S}\mathcal{F}^{-1}(\mathcal{F}(\hat{\ten{K}}[:,:,i,j])\odot \mathcal{F}(\ten{X}[i,:,:])),
\end{equation}
where $\odot$ stands for the point-wise multiplication (also called Hadamard product), and $\hat{\ten{K}} \in \mathbb{R}^{H\times W\times S\times T}$ is the expanded filter, whose slices $\hat{\ten{K}}[:,:,i,j]$ are the torus form of the corresponding slices $\ten{K}[:,:,i,j]$ in the original filter.
\end{theorem}

% some explanation of Theorem 1
By writing Eq.~\ref{eq:freq_conv} alternatively as 
% eq
\begin{equation}
\label{eq:fft_fft}
\mathcal{F}(\ten{Y}[j,:,:]) = \sum_{i=1}^{S} \mathcal{F}(\hat{\ten{K}}[:,:,i,j]) \odot \mathcal{F}(\ten{X}[i,:,:]),
\end{equation}
the input, filter, and the corresponding output are all in the frequency domain. The equivalence between spatial convolution and spectral point-wise product establishes the credibility of using frequency-domain information. Taking a sample image from CIFAR-10 and three filters from the first CONV layer in a pre-trained VGGNet, Figure~\ref{fig:fftkx} visualizes convolution in the frequency domain. More visualization results are available in Appendix 2.1. 

%% in frequency domain
\subsection{Energy Zoning}
\label{sec:HRank_frequency}
% observations in 3D plots
According to the rightmost column row of Figure~\ref{fig:fftkx}, the concentration of low-frequency components of the generated feature map is inversely proportional to the average rank of the feature maps. Specifically, the average rank of the feature maps generated by the filter selected in the first row is the smallest, its low-frequency components of the feature map are the most concentrated. In the third row, the selected filter generates feature maps with high average ranks, and it is obvious that the frequency components of its feature map are much more dispersed. This phenomenon shows that a filter generating feature maps with high average ranks in the spatial domain is expected to generate feature maps with dispersed frequency components. Figure~\ref{fig:rank_energy} uses a toy example to conceptually depict a low-rank channel matrix and a high-rank one. We use zero-valued and nonzero-valued elements to represent high-frequency and low-frequency components, respectively. It is obvious that the right matrix with a high rank has more dispersed nonzero elements\footnote{We remark that the rank of a channel matrix is \textit{invariant} in the spatial and spectral domains as 2D FFT is a bijective linear mapping.}. 

To precisely quantify,  we define a new metric called \textit{energy zone} due to its underlying nature of summing spectral components related to power. The energy zone here refers to an area obtained by expanding outward from the DC center. Due to the conjugate symmetry after applying $\mathcal{F}(\cdot)$ and ${\rm fftshift}(\cdot)$ on the feature map, we define the energy zone with a \textbf{symmetrical} shape. For \textbf{best visualization}, we choose \textbf{circle} as the energy-zone shape, and choose a feature map with a large spatial size.

% observation of the circle plots and motivation to EZCrop
Figure~\ref{fig:rank_in_frequency1} shows the results taking the first CONV in VGGNet having 64 filters as an example, the energy in each circle takes up $70\%$ of the total feature map. The details of how to draw the circles are provided in Appendix 2.1. The 64 subfigures are arranged in an ascending order of the rank-based HRank scheme in~\cite{lin2020hrank} from left to right and top to bottom. In general, we can see that the radius of the circle increases as the average rank of the feature maps increases, which shows the high consistency between the rank-based scheme in the spatial domain and the dispersion of the feature maps' low-frequency components in the frequency domain. However, the circle radii are not monotonically increasing, which means there are still subtle differences in the evaluation of certain filters based on information in the spatial and frequency domains. We remark that analyzing the low-frequency components' concentration brings in \textbf{higher resolution} in channels' importance evaluation since every element in $H_i \times W_i$ is considered. In contrast, the rank of each feature map slice is only an integer in $[0,min(H_i,W_i)]$. This is the reason why HRank and energy metric may produce subtle discrepancy in their channel importance measure. More discussions on the rank-based and energy-based metrics are in Appendix 3. 

To further benchmark the difference in filter importance evaluation, we propose a pruning metric based on the energy zone in Section~\ref{sec:EZCrop} and evaluate it through various experiments in Section~\ref{sec:exp}.

%% fig: rate example
\begin{figure*}[t]
\centering
\includegraphics[scale=0.4]{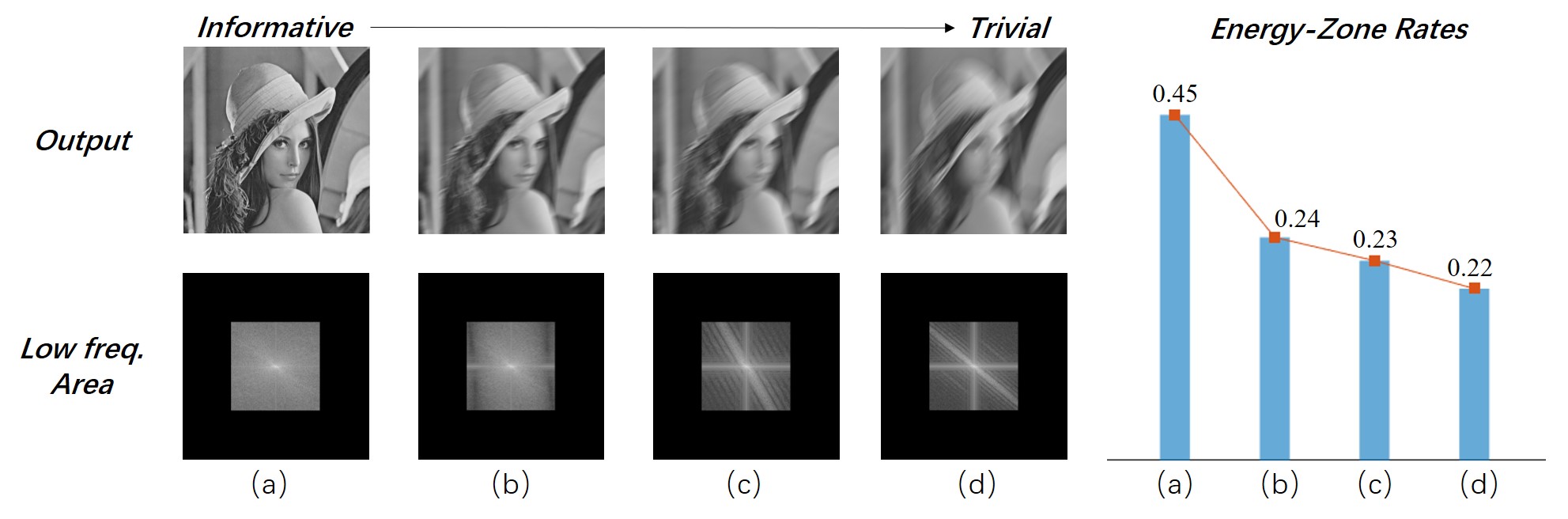}
\caption{By assuming the pictures as the outputs generated by four filters, we give an example to show intuitively how energy-zone ratios reflect the importance of the corresponding filters using the frequency information. (Left) The first row contains four feature map slices, the quality of which decreases from left to right. The second row shows the corresponding selected energy areas. It is worth noting that the concentration of the energy distribution increases as the feature maps become less informative. (Right) The more information the feature map contains, the larger the energy-zone ratio, and the more important the corresponding filter is.}
\label{fig:rate_example}
\end{figure*}

%%%% EZCrop
\section{EZCrop} 
\label{sec:EZCrop}
%% Why choose square but not circle or other shape
Recall from Section~\ref{sec:HRank_frequency} that the radius can be used to evaluate the concentration of the feature map's frequency components. In particular, filters with a large average corresponding radius are treated as important ones. However, this method \textbf{loses resolution} when $H_i$ and $W_i$ of the feature maps $\ten{Y}_i$ are \textbf{small}. For example, it is difficult to determine a proper radius when the feature map slice is of size $4\times 4$ or smaller, such as $2\times 2$. Subsequently, a simple metric based on the concept of energy zone regardless of the feature maps' size is desired. Here, we set the shape of an energy zone to be \textbf{square} and evaluate the importance of filters by four steps:1) find the square center, 2) decide the size of the square, 3) calculate the energy-zone ratios, and 4) sort the filters according to the ratios. These steps are detailed below and the pseudocode can be found in Appendix 4.2.

%% step 1
\textbf{Step 1: Find the Square Center.} For a given feature map $\ten{Y}_i$, as long as $H_i$ and $W_i$ are known, we can determine the coordinate of the DC center by
% eq: find the center
\begin{small}
\begin{subequations}
\label{eq:centers}
% x
\begin{equation}
\begin{aligned}
x_i = \left\{ \begin{array}{lr}
\frac{H_i}{2}+1 ,& H_i \text{ is even}  \vspace{1ex}\\
\frac{H_i+1}{2} ,& H_i \text{ is odd}
\end{array} \right.,
\end{aligned}
\end{equation}
% y
\begin{equation}
y_i = \left\{ \begin{array}{lr}
\frac{W_i}{2}+1 ,& W_i \text{ is even}  \vspace{1ex}\\
\frac{W_i+1}{2} ,& W_i \text{ is odd}
\end{array}\right.,%
\end{equation}%
\end{subequations}%
\end{small}%
regardless of the parity of $H_i$ and $W_i$ or whether $H_i$ and $W_i$ are equal. Figure~\ref{fig:energy_area} visualize the square center when $H_i$ and $W_i$ are equal and are both even, and the schematic diagrams for other scenarios are available in Appendix 4.1.

%% step 2
\textbf{Step 2: Decide the Expanding Distance.} The size of a energy-zoned square is constrained by limiting the distance away from the center, accordingly:
% eq: decide the size of the square
\begin{small}
\begin{equation}
\label{eq:length}
\begin{aligned}
l_{ih} &= H_i - x_i,~l_{iw} = W_i - y_i, \\
d_i &= \begin{cases}
0 , \quad x_i-1 =0 \text{ or } y_i-1=0\\
{\rm ceil}(\beta \cdot \min(l_{ih}, l_{iw})), \quad \text{else} 
\end{cases},%
\end{aligned}%
\end{equation}%
\end{small}%
where $\beta \in (0,1)$ is a hyper-parameter to limit the area. It is worth noting that, $d_i = 0$ means only the DC component itself is contained in the energy-zoned square. Figure~\ref{fig:energy_area} shows the examples when $d_i = 0,1,2$.

%% step 3
\textbf{Step 3: Calculate the Energy Zone Ratios.} Different from Section~\ref{sec:HRank_frequency}, we determine the size of the energy zone first instead of the ratio of the sum of the elements in the energy zone. In this case, we evaluate the concentration of feature map's low-frequency components by the proportion of the sum of elements outside the energy-zoned square:
% eq: calculate the energy-zoned ratio
\begin{small}
\begin{equation}
\label{eq:ratio}
\eta_i^j = \frac{1}{B}\cdot \left( 1-\sum_{b=1}^{B} \frac{S(d_i[b])}{S(\mat{E}_i^j[b,:,:])} \right),%
\end{equation}%
\end{small}%
where $S(d_i[b])$ and {$S(\mat{E}_i^j[b,:,:])$} denote the sum of elements in the energy-zoned square with $d_i$ as the extended length and the sum of elements in the whole square, respectively, for the $b$-th sample in a batch, and $\eta_i^j$ is the energy zone ratio of $j$-th filter in the $i$-th CONV layer, where larger value means more important the filter is. In Figure~\ref{fig:rank_in_frequency2}, we compare the results of the evaluation of the filters' importance through HRank~\cite{lin2020hrank} and EZCrop. Consistent with the conclusion of Section~\ref{sec:HRank_frequency}, the two metrics generally track each other, especially for particularly important or unimportant channels/filters. For filters between the two poles, the evaluation results are slightly different. In Section~\ref{sec:exp}, we use sufficient experiments to show EZCrop indeed provides a better scrutiny of important filters.

%% step 4
\textbf{Step 4: Sort the Filters.} After finishing the above steps, for each CONV layer, we can obtain a set of energy zone ratios $\eta_i = \{ \eta_i^{1}, \eta_i^{2}, \cdots, \eta_i^{T_i}  \}$. The set $I_i = \{ I_i^{1}, I_i^{2}, \cdots, I_i^{T_i} \}$ records the channel index of the elements in $\eta_i$ after sorting them in a descending order. For example, if $\eta_i^{12}$ ranks first in $\eta_i$, then $I_i^{1} = 12$. Figure~\ref{fig:rate_example} shows an example how we sort the filters according to the energy ratios.

%% Complexity Analysis
\textbf{Complexity Analysis.} For a given feature map slice $\mat{Y}_i^j[b,:,:]\in \mathbb{R}^{H_i \times W_i}$, we compare the complexity of HRank and EZCrop to generate the pruning metrics in spatial and frequency domains, respectively. HRank~\cite{lin2020hrank} employs the function $\rm torch.matrix\_rank(\cdot)$ in PyTorch, which uses SVD to compute the matrix rank at a complexity of $\mathcal{O}(\max(H_i, W_i)^3)$. For EZCrop, applying 2D FFT to transform $\mat{Y}_i^j[b,:,:]$ to the frequency domain is the most time-consuming step, which we implemented by $\rm np.fft.fft2(\cdot)$ in NumPy with a complexity of $\mathcal{O}(H_i W_i \log (H_i W_i))$. For a simpler expression, we rewrite it as $\mathcal{O}(n^2 \log n)$ which indicates the high efficiency of EZCrop versus that of HRank. 

%%%% Experiments
\section{Experimental Results}
\label{sec:exp}

%%% configuration
We conduct extensive experiments on image classification tasks, using CIFAR-10~\cite{krizhevsky2009learning} and ImageNet~\cite{deng2009imagenet} datasets, to demonstrate the superiority of the proposed EZCrop channel pruning scheme. The CNN models under test are all popular ones, namely, VGGNet~\cite{simonyan2014very}, ResNet~\cite{he2016deep}, and DenseNet~\cite{huang2017densely}. We present the performance of the state-of-the-art algorithms for a comprehensive overview, and we mainly compare EZCrop against its closest scheme HRank which uses channel matrix rank in the spatial domain to measure channel importance. 

The compared pruning results shown in Tables~\ref{tab:cifar10_vgg16} to~\ref{tab:imagenet_resnet} are directly acquired from the corresponding papers or the official GitHub repository. We get the result of multi-pass HRank in Table~\ref{tab:repetitive_resnet} ourselves by generalizing the single-pass scheme, since the authors of~\cite{lin2020hrank} did not report the results of applying HRank repetitively. For the sake of simplicity, We fix $\beta=0.25$ in Eq.~\ref{eq:length} in all experiments when generating the energy zone ratios, more training details can be found in Appendix 5. All experiments are run on a computer equipped with four NVIDIA GeForce GTX1080Ti Graphics Cards, each with 11GB frame buffer. 

% Time
\textbf{Time Comparison.} In Table~\ref{tab:time}, we further show the actual runtimes of EZCrop and HRank to generate the required pruning metrics for those CIFAR-10/ImageNet networks on one GTX1080Ti GPU. It is obvious that the time required for EZCrop to generate pruning metric is much shorter in all models. Noticeably, the larger the dataset, the more pronounced the efficiency of EZCrop is.

% tab: Time
\begin{table}[t]
\scriptsize
\centering
\setlength{\tabcolsep}{3.5mm}{
\begin{tabular}{ccc}
\toprule
~ & HRrank~\cite{lin2020hrank} & \textbf{EZCrop} ($\downarrow$) \\
\midrule
VGGNet & $1505.54$s & $\mathbf{356.94}$s $(76.29\%)$ \\
ResNet-56 & $1247.51$s & $\mathbf{381.97}$s $(69.38\%)$ \\
DenseNet-40 & $473.17$s & $\mathbf{171.50}$s $(63.76\%)$\\
ResNet-50 & $7.96$h & $\mathbf{3.45}$h $(56.6\%)$ \\
\bottomrule
\end{tabular}
}
\caption{\centering{Runtimes for metric computation in HRank and EZCrop on CIFAR-10 / ImageNet (bottom row) nets.}}
\label{tab:time}
\end{table}

%% cifar-10
\subsection{CIFAR-10}
\label{sec:exp_cifar10}

% VGGNet
\textbf{VGGNet.} The VGG-16 used in the experiments is a variation of the original VGGNet for CIFAR-10 dataset taken from~\cite{li2016pruning}. In Table~\ref{tab:cifar10_vgg16}, the upper part is the performances of the state-of-the-art pruning approaches excluding HRank, while the lower part focuses on the comparison between HRank and EZCrop. The tables in the following context are in the same format. First, we compare EZCrop with the upper part approaches. Compared with FPGM, though both of them can reach the accuracy of around $94\%$, EZCrop has higher computational efficiency with more significant FLOPs reduction ($58.1\%$ vs. $35.9\%$). When compared with L1, SSS, Zhao \textit{et al.} , GAL, and Wang \textit{et al.}, EZCrop has advantages in all aspects that it obtains the fastest and most compact model with the best performance. Furthermore, when compared with HRank under the same settings, namely, same FLOPs and Params reductions, EZCrop outperforms HRank in both cases ($94.01\%$ vs. $93.73\%$ / $93.70\%$ vs. $93.56\%$). This reveals that though HRank can evaluate the filters well using the information in the spatial domain, EZCrop offers more accurate evaluation benefiting from the additional frequency information. 

% tab: Pruning results of VGGNet on CIFAR-10
\begin{table}[t]
\scriptsize
\setlength{\tabcolsep}{0.8mm}
\centering
\begin{tabular}{cccc}
\toprule
Model & Top-1\% & FLOPs ($\downarrow$) & Params ($\downarrow$) \\
\midrule
VGGNet & $93.96$ & $313.73$M$(0.0\%)$ & $14.98$M$(0.0\%)$\\
L1~\cite{li2016pruning} & $93.40$ & $206.00$M$(34.3\%)$ & $5.40$M$(64.0\%)$\\
SSS~\cite{huang2018data} & $93.02$ & $183.13$M$(41.6\%)$ & $3.93$M$(73.8\%)$\\
Zhao \textit{et al.}~\cite{zhao2019variational} & $93.18$ & $190.00$M$(39.1\%)$ & $3.92$M$(73.3\%)$ \\
GAL-$0.05$~\cite{lin2019towards} & $92.03$ & $189.49$M$(39.6\%)$ & $3.36$M$(77.6\%)$ \\
GAL-$0.1$~\cite{lin2019towards} & $90.78$ & $171.89$M$(45.2\%)$ & $2.67$M$(82.2\%)$\\
FPGM~\cite{he2019filter} & $94.00$ & $201.10$M$(35.9\%)$ & $-$\\
Wang\textit{et al.}~\cite{wang2020pruning} & $93.63$ & $156.86$M$(50.0\%)$ & $-$\\
\midrule
\midrule
HRank~\cite{lin2020hrank} & $93.73$ & $131.17$M$(58.1\%)$ & $2.76$M$(81.6\%)$ \\
\textbf{EZCrop} & $\mathbf{94.01}$ & $\mathbf{131.17}$M$(58.1\%)$ & $\mathbf{2.76}$M$(81.6\%)$\\
HRank~\cite{lin2020hrank} & $93.56$ & $104.78$M$(66.6\%)$ & $2.50$M$(83.3\%)$\\
\textbf{EZCrop} & $\mathbf{93.70}$ & $\mathbf{104.78}$M$(66.6\%)$ & $\mathbf{2.50}$M$(83.3\%)$ \\
\bottomrule
\end{tabular}
\caption{\centering{Pruning results of VGGNet on CIFAR-10.}}
\label{tab:cifar10_vgg16}

%%% ResNet-56
\vspace{5mm}
\scriptsize
\centering
\setlength{\tabcolsep}{1.0mm}
\begin{tabular}{cccc}
\toprule
Model & Top-1\% & FLOPs ($\downarrow$) & Params ($\downarrow$) \\
\midrule
ResNet-56 & $93.26$ & $125.49$M$(0.0\%)$ & $0.85$M$(0.0\%)$\\
L1~\cite{li2016pruning} & $93.06$ & $90.90$M$(27.6\%)$ & $0.73$M$(14.1\%)$\\
He \textit{et al.}~\cite{he2017channel} & $90.80$ & $62.00$M$(50.6\%)$ & $-$\\
NISP~\cite{yu2018nisp} & $93.01$ & $81.00$M$(35.5\%)$ & $0.49$M$(42.4\%)$\\
GAL-0.6~\cite{lin2019towards} & $92.98$ & $78.30$M$(37.6\%)$ & $0.75$M$(11.8\%)$\\
GAL-0.8~\cite{lin2019towards} & $90.36$ & $49.44$M$(60.2\%)$ & $0.29$M$(65.9\%)$\\
FPGM~\cite{he2019filter} & $93.49$ & $59.44$M$(52.6\%)$ & $-$\\
Wang \textit{et al.}~\cite{wang2020pruning} & $93.05$ & $62.75$M$(50.0\%)$ & $-$ \\
\midrule
\midrule
HRank~\cite{lin2020hrank} & $93.85$ & $90.35$M$(28.0\%)$ & $0.66$M$(22.3\%)$ \\
\textbf{EZCrop} & $\mathbf{94.07}$ & $\mathbf{90.35}$M$(28.0\%)$ & $\mathbf{0.66}$M$(22.3\%)$\\
HRank~\cite{lin2020hrank} & $93.57$ & $65.94$M$(47.4\%)$ & $0.48$M$(42.8\%)$\\
\textbf{EZCrop} & $\mathbf{93.80}$ & $\mathbf{65.94}$M$(47.4\%)$ & $\mathbf{0.48}$M$(42.8\%)$\\
HRank~\cite{lin2020hrank} & $92.32$ & $34.78$M$(74.1\%)$ & $0.24$M$(70.0\%)$\\
\textbf{EZCrop} & $\mathbf{92.52}$ & $\mathbf{34.78}$M$(74.1\%)$ & $\mathbf{0.24}$M$(70.0\%)$ \\
\bottomrule
\end{tabular}
\caption{\centering{Pruning results of ResNet-56 on CIFAR-10.}}
\label{tab:resnet-56}
\end{table}

% tab: Pruning results of DenseNet-40 on CIFAR-10
\begin{table}[t]
\scriptsize
\centering
\setlength{\tabcolsep}{0.6mm}{
\begin{tabular}{cccc}
\toprule
Model & Top-1\% & FLOPs ($\downarrow$) & Params ($\downarrow$)\\
\midrule
DenseNet-40 & $94.82$ & $282.00$M$(0.0\%)$ & $1.04$M$(0.0\%)$\\
Liu \textit{et al.}-40\%~\cite{liu2017learning} & $\mathbf{94.81}$ & $190.00$M$(32.8\%)$ & $0.66$M$(36.5\%)$\\
GAL-0.01~\cite{lin2019towards} & $94.29$ & $182.92$M$(35.3\%)$ & $0.67$M$(35.6\%)$\\
GAL-0.05~\cite{lin2019towards} & $93.53$ & $128.11$M$(54.7\%)$ & $0.45$M$(56.7\%)$\\
Zhao \textit{et al.}~\cite{zhao2019variational} & $93.16$ & $156.00$M$(44.8\%)$ & $0.42$M$(59.7\%)$\\
\midrule
\midrule
HRank~\cite{lin2020hrank} & $94.51$ & $173.39$M$(38.5\%)$ & $0.62$M$(40.1\%)$\\
\textbf{EZCrop} & $\mathbf{94.72}$ & $\mathbf{173.39}$M$(38.5\%)$ & $\mathbf{0.62}$M$(40.1\%)$\\
HRank~\cite{lin2020hrank} & $93.66$ & $113.08$M$(59.9\%)$ & $0.39$M$(61.9\%)$\\
\textbf{EZCrop} & $\mathbf{93.76}$ & $\mathbf{113.08}$M$(59.9\%)$ & $\mathbf{0.39}$M$(61.9\%)$\\
\bottomrule
\end{tabular}}
\caption{\centering{Pruning results of DenseNet-40 on CIFAR-10.}}
\label{tab:cifar10_densenet}
\end{table}

% ResNet-56 
\textbf{ResNet-56.} Table~\ref{tab:resnet-56} shows the results of different approaches on ResNet-56. Compared with L1, when their FLOPs reductions are both around $30\%$, EZCrop discards more parameters $(22.3\%)$ than L1 $(14.1\%)$, and achieves better performance ($94.07\%$ vs. $93.06\%$). Compared with He \textit{et al.}, FPGM and Wang, when their FLOPs reductions are both around $50\%$, EZCrop has the highest accuracy of $93.80\%$. When compared with GAL, EZCrop outperforms it in all aspects. The following are three sets of comparative experiments between EZCrop and HRank, where EZCrop achieves higher accuracy for all settings ($94.07\%$ vs. $93.85\%$ , $93.80\%$ vs. $93.57\%$ and $92.52\%$ vs. $92.32\%$ ). This shows EZCrop not only outperforms HRank on networks with plain structures but also residual blocks.

% DenseNet-40
\textbf{DenseNet-40.} The results of DenseNet-40 are given in Table~\ref{tab:cifar10_densenet}. Compared Liu \textit{et al.}, EZCrop achieves higher compression rate (Params: $0.62$M vs. $0.66$M) and faster acceleration (FLOPs: $173.39$M vs. $190.00$M) with only $0.09\%$ lower accuracy. This shows EZCrop is more promising to achieve better performance when highly compact models are required. Compared with GAL-0.01, EZCrop has advantages in all aspects. As for GAL-0.05 and Zhao \textit{et al.}, when the Params reductions are all around $60\%$, EZCrop obtains the best performance (Top-$1\%$: $93.76\%$) with the highest FLOPs reduction ($59.9\%$). In the two comparative experiments with HRank, EZCrop achieves higher Top-1 accuracy in both cases, namely, $94.72\%$ vs. $94.51\%$ and $93.76\%$ vs. $93.66\%$. This observation verifies that EZCrop beats HRank also on networks with inception modules.

%%%% ImageNet
\subsection{ImageNet}
\label{sec:exp_imagenet}
\textbf{ResNet-50.} The results on ImageNet, a more complicated dataset than CIFAR-10, are displayed in Table~\ref{tab:imagenet_resnet}. When the number of parameters is around $15$M, EZCrop ($75.68\%$) has higher accuracy than SSS-26 ($71.82\%$) and GAL-1 ($69.31\%$). Though GAL-1 can reduce the FLOPs more significantly than EZCrop (FLOPs: $1.58$B vs. $2.26$B), its accuracy drops drastically at the same time, which cannot be ignored. When the number of parameters is around $10$M, EZCrop outperforms ThiNet-50 and GAL-1-joint with significant advantages on the performance ($74.33\%$ vs. $66.42\%$ / $69.31\%$). For the remaining upper part approaches, EZCrop has advantages in all aspects. For the two comparative experiments with HRank, EZCrop boosts the accuracy in both cases. This indicates that EZCrop can deal with large and complicated datasets as well.

% tab: Pruning Results of ResNet-50 on ImageNet
\begin{table}[t]
\centering
\scriptsize
\setlength{\tabcolsep}{1.4mm}
\begin{tabular}{ccccc}
\toprule
Model & Top-1\% & Top-5\% & FLOPs & Params \\
\midrule
ResNet-50~\cite{luo2017thinet} & $76.15$ & $92.87$ & $4.09$B & $25.50$M\\
He \textit{et al.}~\cite{he2017channel} & $72.30$ & $90.80$ & $2.73$B & $-$\\
ThiNet-50~\cite{luo2017thinet} & $68.42$ & $88.30$ & $1.10$B & $8.66$M\\
SSS-26~\cite{huang2018data} & $71.82$ & $90.79$ & $2.33$B & $15.60$M\\
SSS-32~\cite{huang2018data} & $74.18$ & $91.91$ & $2.82$B & $18.60$M\\
GDP-0.5~\cite{lin2018accelerating} & $69.58$ & $90.14$ & $1.57$B & $-$\\
GDP-0.6~\cite{lin2018accelerating} & $71.19$ & $90.71$ & $1.88$B & $-$\\
GAL-0.5~\cite{lin2019towards} & $71.95$ & $90.94$ & $2.33$B & $21.20$M\\
GAL-1~\cite{lin2019towards} & $69.88$ & $89.75$ & $1.58$B & $14.67$M\\
GAL-0.5-joint~\cite{lin2019towards} & $71.80$ & $90.82$ & $1.84$B & $19.31$M\\
GAL-1-joint~\cite{lin2019towards} & $69.31$ & $89.12$ & $1.11$B & $10.21$M\\
FPGM~\cite{he2019filter} & $75.91$ & $92.63$ & $2.36$B & $-$ \\
\midrule
\midrule
HRank~\cite{lin2020hrank} & $75.56$ & $92.63$ & $2.26$B & $15.09$M\\
\textbf{EZCrop} & $\textbf{75.68}$ & $\textbf{92.70}$ & $2.26$B & $15.09$M\\
HRank~\cite{lin2020hrank} & $74.19$ & $91.94$ & $1.52$B & $11.05$M\\
\textbf{EZCrop} & $\textbf{74.33}$ & $\textbf{92.00}$ & $1.52$B & $11.05$M\\
\bottomrule
\end{tabular}
\caption{\centering{Pruning Results of ResNet-50 on ImageNet.}}
\label{tab:imagenet_resnet}
\end{table}

% tab: repetitive pruning on ResNet-56
\begin{table}[t]
\centering
\scriptsize
\setlength{\tabcolsep}{1.1mm}{
\begin{tabular}{ccccc}
\toprule
\#Passes (\#epochs) & FLOPs & Params & HRank~\cite{lin2020hrank} & \textbf{EZCrop}\\
\midrule
1 ($300$) & $90.86M$ & $0.63M$& 93.76\% & \textbf{93.95}\% \\
2 ($300$) & $66.25M$ & $0.46M$& 93.15\% & \textbf{93.42}\% \\
3 ($300$) & $36.03M$ & $0.22M$& 91.58\% & \textbf{92.18}\% \\
\bottomrule
\end{tabular}
}
\caption{\centering{Repetitive pruning of ResNet-56 on CIFAR-10.}}
\label{tab:repetitive_resnet}
% \end{table}
\vspace{5mm}
% %%%% tab: std
% \begin{table}[t]
\scriptsize
\centering
\setlength{\tabcolsep}{1.2mm}{
\begin{tabular}{cccc}
\toprule
Mean Acc. (\%) / STD  & VGGNet & ResNet-56 & DenseNet-40 \\
\midrule
EZCrop  & $\mathbf{93.98 / 0.073}$ & $\mathbf{93.99 / 0.097}$ & $\mathbf{94.63 / 0.066}$ \\
HRank~\cite{lin2020hrank} & $93.75 / 0.140$ & $93.76 / 0.168$ & $94.26 / 0.165$ \\
\bottomrule
\end{tabular}
}
\caption{The mean and STD of the accuracy by conducting the experiments on CIFAR-10 for $20$ times on VGGNet / ResNet-56 / DenseNet-40.}
\label{tab:std}
\end{table}

% repetitive pruning
\subsection{Robustness Under Repetitive Pruning}
\label{sec:exp_repetitive_pruning}
In this section, we test both HRank and EZCrop in a multi-pass manner, i.e., pruning models repetitively. Repetitive pruning requires multiple evaluations of channels' importance, which amplifies the difference between the two schemes. We conduct the experiments using ResNet-56 on CIFAR-10, and keep adopting HRank and EZCrop to get a more compact model without a fixed compression rate, then compare their performance. For fairness, all settings of HRank and EZCrop for every single-pass are the same. Table~\ref{tab:repetitive_resnet} shows that when getting more compact models, the gap of EZCrop over HRank is enlarged after each single-pass, i.e., EZCrop enjoys higher robustness under repetitive pruning, and is able to evaluate the filters more accurately in each single-pass.

%% std
\subsection{Standard Deviation (STD) Analysis}
\label{sec:std}
To further show the effectiveness and robustness of EZCrop while dispelling the worries that the improvements compared with HRank are due to standard variations, we conduct the first comparative experiments in Table~\ref{tab:cifar10_vgg16}-\ref{tab:cifar10_densenet} for $20$ times and compute the average and STD of the accuracy, respectively. The results in Table~\ref{fig:ablation1} prove that the improvements are not accidental, and EZCrop actually makes more accurate and robust evaluations of the importance of filters.

%%%% ablation study
\subsection{Ablation Study} 
\label{sec:abliation}
Excluding the pruning metrics, the compression rate and training settings for VGGNet / ResNet-56 / DenseNet-40 in this section are the same as the first comparative experiments in Table~\ref{tab:cifar10_vgg16}-\ref{tab:cifar10_densenet}, respectively.
%% Pruning with low energy rate or randomly
\subsubsection{Effectiveness of Energy-zone Rate}
\label{sec:rate_efficient}
\begin{figure}[t]
\centering
\includegraphics[scale=0.25]{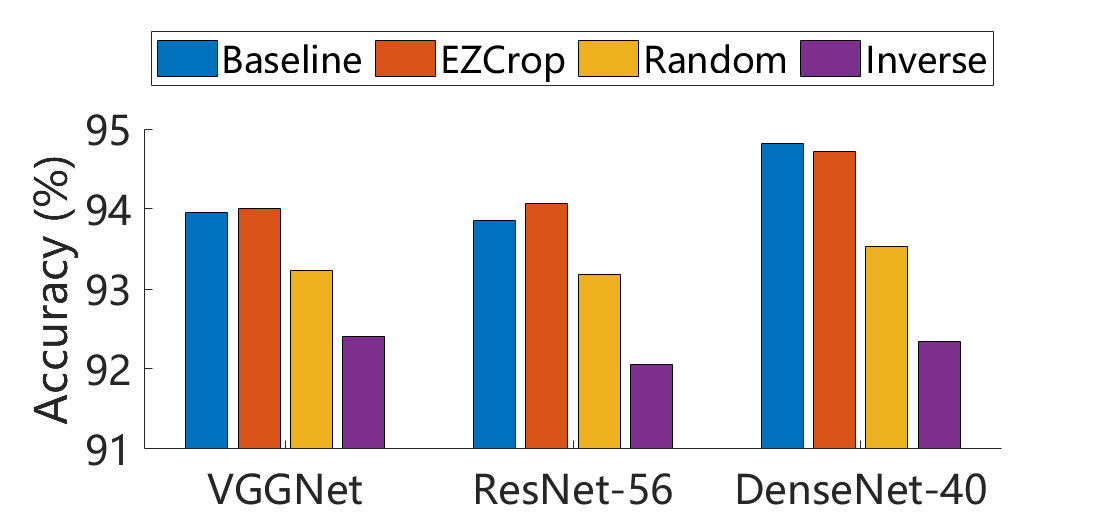}
\caption{Performances of different pruning approaches when pruning VGGNet / ResNet-56 / DenseNet-40 on CIFAR-10. }
\label{fig:ablation1}
\end{figure}

To show that EZCrop makes reliable filter importance evaluation, we compare the results by pruning the models: 1) using EZCrop, 2) randomly, 3) using inverse EZCrop (filters with high energy-zone ratio will be removed). For each network in Figure~\ref{fig:ablation1}, we observe that EZCrop outperforms random pruning, while the inverse EZCrop has the worst performance. The gaps in the accuracy confirm the effectiveness of our energy-zone metric in the frequency domain.

% fig: ablation 1

% fig
\begin{figure}[t]
\centering
\includegraphics[scale=0.27]{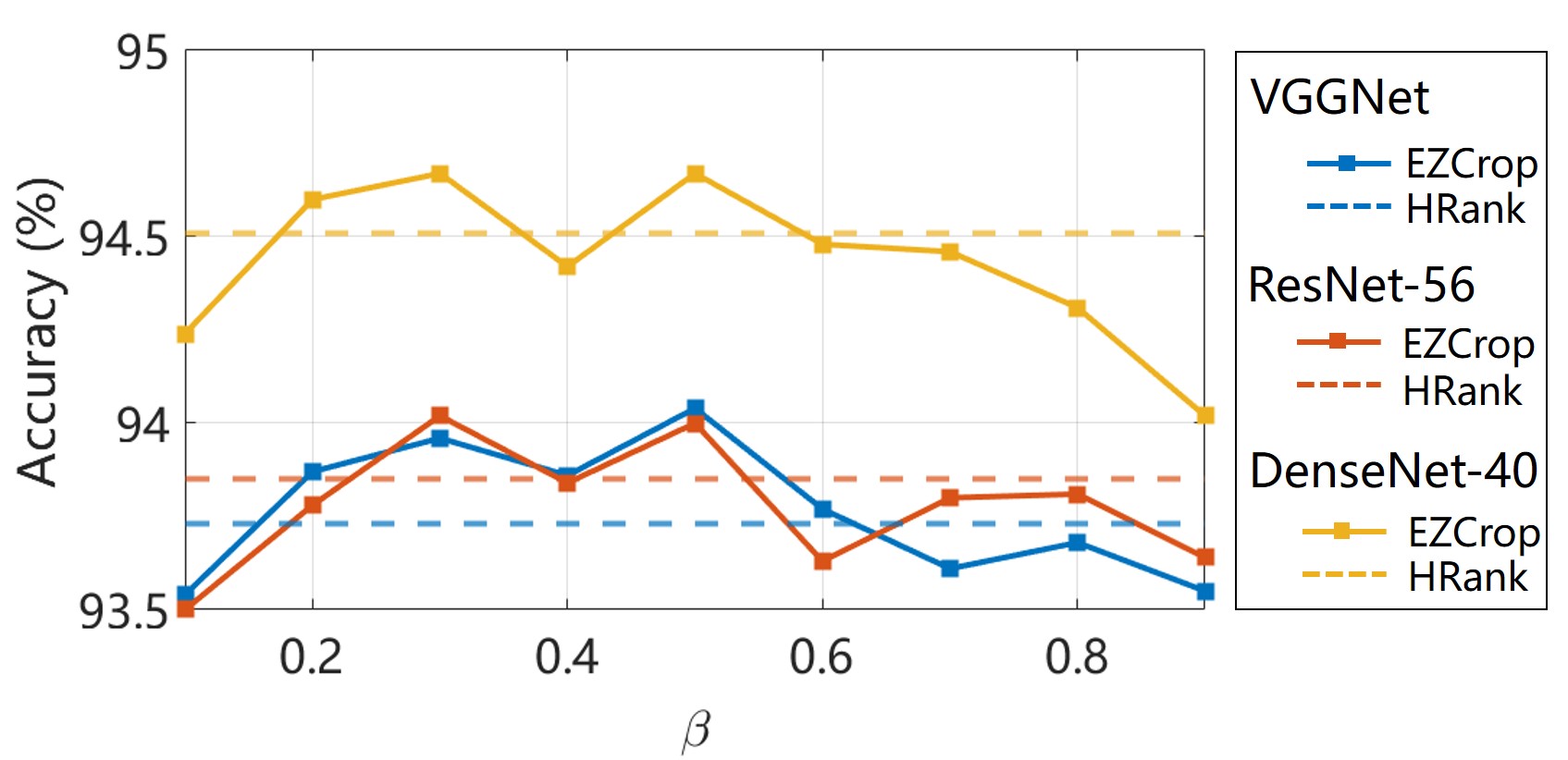}
\caption{Relations between accuracy and $\beta$, which decides the expanding distance. The larger $\beta$, the larger the energy zone.}
\label{fig:step}
\end{figure}

%% Pruning with different steps
\subsubsection{Appropriate Expending Distance}
\label{sec:study_step}
There is a hyper-parameter $\beta$ in EZCrop when we decide the expanding distance. In this paper, we set $\beta=0.25$ for all experiments for the sake of simplicity. However, this general setting may not be an optimal setting for all. In Figure~\ref{fig:step}, we explore the relations between the accuracy and $\beta$. It is worth noting that the three curves have similar trends, the curve shows an ``M'' shape when $\beta$ is between $0.2$ and $0.6$. Based on this observation, we suggest to manually set $\beta$ around $0.3$ or $0.5$, namely, two peaks of the curve.

%%%% Conclusion
\section{Conclusion}
\label{sec:conclusion}
This work has connected the previously mysterious constant-rank phenomenon in a CNN feature map channel to a novel, analytical view in the frequency domain. Via a spectral perspective, an efficient FFT-based energy-zone metric has been proposed for evaluating the importance of a channel and its corresponding filter. This leads to a channel pruning scheme named EZCrop (energy-zoned channels for robust output pruning) which readily outperforms existing state-of-the-art in terms of compression rates, runtimes and FLOPs. EZCrop also constitutes a robust way for repetitive channel pruning to stretch the limit of CNN compression, while maintaining high output accuracy.%

%%%% Reference
\clearpage
\normalem
\bibliographystyle{icml2021}
\bibliography{egbib}

\end{document}